\documentclass[letterpaper]{llncs}
\usepackage{times}
\usepackage{epsfig}
\usepackage{amssymb}
\usepackage{amsmath}
\usepackage{amsfonts}
\usepackage{graphicx}
\usepackage{subfigure}
\usepackage{verbatim}
\usepackage{setspace}

\newcommand{\reshape}{ReSHAPE}

\mainmatter
\title{\bf Efficient Multidimensional Data Redistribution for Resizable Parallel Computations\\ 
}

\author{Rajesh Sudarsan and Calvin J. Ribbens}
\institute{Department of Computer Science\\
Virginia Tech, Blacksburg, VA 24061-0106 \\
\email{\{sudarsar, ribbens\}@vt.edu}}

\begin{document}

\maketitle
\thispagestyle{empty}
\begin{abstract}
Traditional parallel schedulers running on cluster supercomputers
support only static scheduling, where the number of processors
allocated to an application remains fixed throughout the  execution of the job.
This results in under-utilization of idle system resources 
thereby decreasing overall system throughput.
In our research, we have developed a
prototype framework called ReSHAPE, which supports dynamic resizing of
parallel MPI applications executing on distributed memory platforms.
The resizing library in ReSHAPE includes support for releasing and acquiring
processors and efficiently redistributing application state to a new
set of processors. In this paper, we derive an algorithm for redistributing
two-dimensional block-cyclic arrays from $P$ to $Q$ processors, organized
as 2-D processor grids. The algorithm ensures
a contention-free communication schedule for data redistribution
if $P_r \leq Q_r$ and $P_c \leq Q_c$.
In other cases, the algorithm implements
circular row and column shifts on the communication schedule
 to minimize node contention.
\end{abstract}
\keywordname{ Dynamic scheduling, Dynamic resizing, Data redistribution, Dynamic resource management, process remapping, resizable applications}

\section{Introduction}
As terascale supercomputers become more common and as the high-performance computing (HPC) community turns its attention to petascale machines, 
the challenge of providing effective resource management for high-end machines grows in both importance and difficulty. 
A fundamental problem is that conventional parallel schedulers are static, i.e., 
 once a job is allocated a set of resources, they remain fixed throughout the 
life of an application's execution.
It is worth asking whether a dynamic resource manager, which has the ability to 
modify resources allocated to jobs at runtime, would allow more effective resource 
management. The focus of our research is on dynamically reconfiguring parallel 
applications to use a different number of processes, i.e., on 
\textit{dynamic resizing} of applications.
\footnote{A shorter version of this paper is available in the proceedings of the \textit{The Fifth International Symposium on Parallel and Distributed Processing and Applications (ISPA07)}}

In order to explore the potential benefits and challenges of dynamic resizing, we are developing ReSHAPE, a framework for dynamic {\bf Re}sizing and {\bf S}cheduling of {\bf H}omo\-geneous {\bf A}pplications in a {\bf P}arallel {\bf E}nvironment. The ReSHAPE framework includes a programming model and an API, data redistribution algorithms and a runtime library, and a parallel scheduling and resource management system 
framework. 
ReSHAPE allows the number of processors allocated to a parallel message-passing application to be changed at run time.  
It targets long-running iterative computations, i.e., homogeneous computations that perform similar computational steps over and over again. 
By monitoring the performance of such computations on various processor sizes, 
the ReSHAPE scheduler can take advantage of idle processors on large 
clusters to improve the turn-around time of high-priority jobs, 
or shrink low-priority jobs to meet quality-of-service or advanced 
reservation commitments.

\vspace{-0.01in}
Dynamic resizing necessiates runtime application data redistribution.
Many high performance computing applications and mathematical libraries like ScaLAPACK~\cite{scalapack} require block-cyclic data redistribution to achieve 
computational efficiency. 
Data redistribution involves four main stages --- data identification and 
index computation, communication schedule generation, message packing and unpacking
 and finally, data transfer. 
Each processor identifies its part of the data to redistribute and transfers the 
data in the message passing step according to the order specified in the 
communication schedule. A node contention occurs when one or more processors
sends messages to a single processor. A redistribution \textit{communication schedule}
 aims to 
minimize these node contentions and maximiz network bandwidth utilization.
Data is packed or marshalled on the source processor to form a message and is unmarshalled on the destination processor.

\vspace{-0.01in}
In this paper, we present an algorithm for redistributing two-dimensional 
block-cyclic data from $P$ ( $P_r$ rows $\times P_c$ columns) to $Q$ 
($Q_r$ rows $\times Q_c$ columns) processors, organized as 2-D processor grids. 
We evaluate the algorithm's 
performance by measuring the redistribution time for different block-cyclic matrices.
If $P_r \leq Q_r$ and $P_c \leq Q_c$, the algorithm  ensures a contention-free communication schedule for redistributing data from source processor set $P$ to $Q$ processor set.
In other cases the algorithm minimizes node contentions by performing row or column circular shifts on the communication schedule. 
The algorithm discussed in this paper supports 2-D block cyclic data redistribution 
for only one- and two-dimensional processor topology.
We also discuss in detail the modifications needed to port an existing scientific application to use the dynamic resizing capability of ReSHAPE using the API provided by the framework.

The rest of the paper is organized as follows: Section~\ref{sec:relatedworks} discusses prior work in the area of data redistribution. 
Section~\ref{sec:systemoverview} briefly reviews the architecture of the ReSHAPE framework and discusses in detail the two-dimensional redistribution algorithm and the ReSHAPE API. 
Section~\ref{sec:experiments} reports our experimental results of the redistribution algorithm with the ReSHAPE framework tested on the SystemX cluster at Virginia Tech. 
We conclude in Section~\ref{sec:discussion} discussing future directions to this research.

\section{RelatedWork}
\label{sec:relatedworks}
Data redistribution within a cluster using message passing approach has been extensively studied in literature.
Many of the past research efforts~\cite{chung} \cite{desprez98scheduling} \cite{Guo}
\cite{Hsu} \cite{kalns} \cite{kaushik} \cite{Lim96}
 \cite{ramaswamy} \cite{thakur2} \cite{thakur1} \cite{walker} were targeted towards redistributing
cyclically distributed one dimensional arrays between the same set of processors within a cluster on a 1-D processor topology.
To reduce the redistribution overhead cost, Walker and Otto~\cite{walker} and Kaushik~\cite{kaushik} proposed a K-step communication 
schedule based on modulo arithmetic and tensor products repectively.
Ramaswamy and Banerjee~\cite{ramaswamy} proposed a redistribution technique,
PITFALLS, that uses line segments to map array elements to a processor.
This algorithm can handle any arbitrary number of source and destination processors.
However, this algorithm does not use communication schedules during
redistribution resulting in node contentions during data transfer.
Thakur et al.~\cite{thakur1}\cite{thakur2} use  \textit{gcd} and \textit{lcm} methods for redistributing cyclically distributed one dimensional arrays
on the same processor set.
The algorithms described
by Thakur et al.~\cite{thakur2} and Ramaswamy~\cite{ramaswamy} use a series of one-dimensional
redistributions  to handle  multidimensional arrays. This approach can result in
 significant redistribution overhead cost due to unwanted communication.
Kalns and Ni~\cite{kalns} presented a technique for mapping data to processors by
assigning logical processor ranks to the target processors. This technique reduces
 the total amount of data that must be communicated during redistribution.
Hsu et al.~\cite{Hsu}
further extended this work and proposed a generalized processor mapping
technique for redistributing data from cyclic(kx) to cyclic(x), and vice versa.
Here, x denotes the number of data blocks assigned to each processor.
However, this method is applicable only when the number of source and target processors are same.
Chung et al.~\cite{chung} proposed an efficient method for index computation using
basic-cycle calculation (BCC) technique for redistributing data from cyclic(x) to
cyclic(y) on the same processor set. An extension of this work by Hsu et al.~\cite{chung2000}
uses generalized basic-cyclic calculation method to redistribute data
from cyclic(x) over P processors to cyclic(y) over Q processors. The generalized
BCC uses uses bipartite matching approach for data redistribution.
Lim et al.~\cite{Lim96} developed a redistribution framework that could redistribute one-dimensional array from one block-cyclic scheme to another on the same processor set using a generalized circulant matrix formalism.
Their algorithm applies row and column transformations on the communication schedule matrix to generate a conflict-free schedule.

\vspace{-0.02in}
Prylli et al.~\cite{prylli}, Desprez et al.~\cite{desprez98scheduling} and
Lim et al.~\cite{Lim97}
proposed efficient algorithms for redistributing one- and two-dimensional block cyclic
arrays.
Prylli et al.~\cite{prylli} proposed a simple scheduling algorithm, called Caterpillar,
for redistributing data across a two-dimensional processor grid.
At each step $d$ in the algorithm, processor ${P_i} (0 < i \leq P)$ in the destination
processor set exchanges its data
with processor $P_{((P-i-d)\ mod\ P)}$.
The Caterpillar algorithm does not have a global knowledge of the communication schedule
and redistributes the data using the local knowledge of the communications at every step.
As a result, this algorithm is not efficient for data redistribution using ``non-all-to-all''
communication. Also, the redistribution time for a step is the time taken to transfer the
largest message in that step.
Desprez et al.~\cite{desprez98scheduling} proposed a general solution for redistributing
one-dimensional block-cyclic data from a cyclic(x) distribution
on a P-processor grid to a cyclic(y) distribution on a Q-processor grid for arbitrary values of P, Q, x, and y.
The algorithm assumes the source and target processors as
disjoint sets and uses a bipartite matching to compute the communication schedule. However, this algorithm does not ensure a contention-free communication schedule.
In a recent work, Guo and Pan~\cite{Guo} described a method to construct schedules that minimizes number of communication steps, avoids node contentions, and minimizes the effect of difference in message length in each communication step.
Their algorithm focuses on redistributing one-dimensional data from a cyclic(kx) distribution on P processors to cyclic(x) distribution on Q
processors for any arbitrary positive values of P and Q. 
Lim et al.~\cite{Lim97} propose an algorithm for
redistributing a two-dimensional block-cyclic array across a two-dimensional processor grid. But the algorithm  is restricted to redistributing data across
different processor topologies on the same processor set.
Park~et~al.~\cite{park} extended the idea described by Lim et al.~\cite{Lim97} and proposed an algorithm for redistributing one-dimensional block-cyclic array with cyclic(x)
distribution on P processors to cyclic(kx) on Q processors where P and Q  can be any arbitrary positive value.

To summarize, most of the existing approaches either deal with redistribution of
block-cyclic array across one-dimensional processor topology on the same or
on a different processor set. The Caterpillar algorithm by Prylli et al.~\cite{prylli}
is the closest related work to our redistribution algorithm in that it supports
redistribution on checkerboard processor topology.
In our work,
we extend the idea in~\cite{Lim97}\cite{park} to develop an algorithm to redistribute two-dimensional block-cyclic data
distributed across a 2-D processor grid topology.
The data is redistributed from $P$ ($P_r \times P_c$) to Q ($Q_r \times Q_c$) processors where P and Q can be any  arbitrary positive value.
Our work is contrary to  Desprez et al.~\cite{desprez98scheduling} where they assume that there is no overlap among processors in the
source and destination processor set. Our algorithm builds an efficient communication schedule and uses non-all-to-all communication
for data redistribution. 
We apply row and column transformations using the
circulant matrix formalism to minimize
node contentions in the communication schedule.

\section{System Overview}
\label{sec:systemoverview}
\begin{figure}[t]
\begin{center}
\subfigure[]{
\includegraphics[scale=0.28]{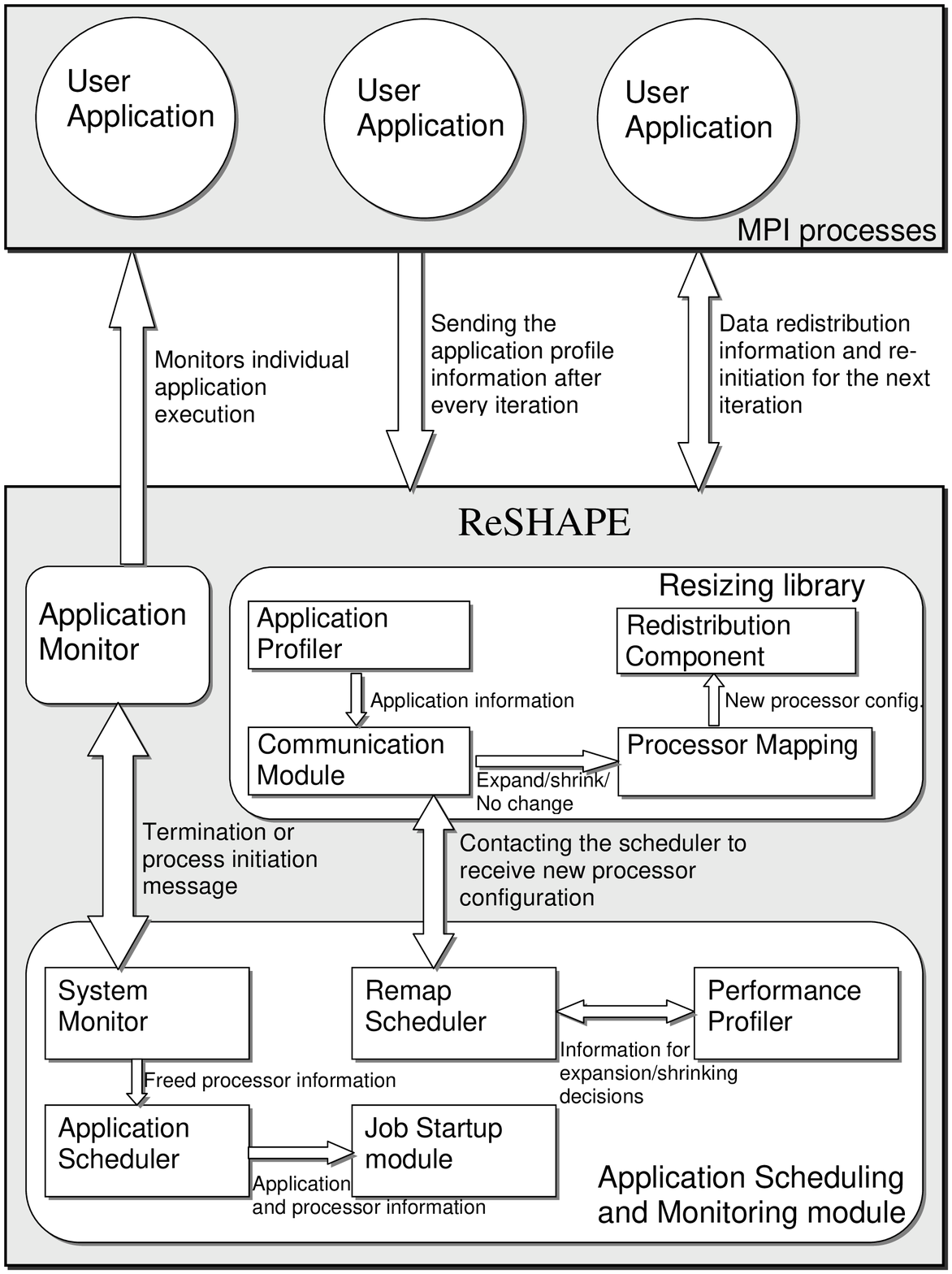}
\label{fig:design}
}
\subfigure[]{
\includegraphics[scale=0.28]{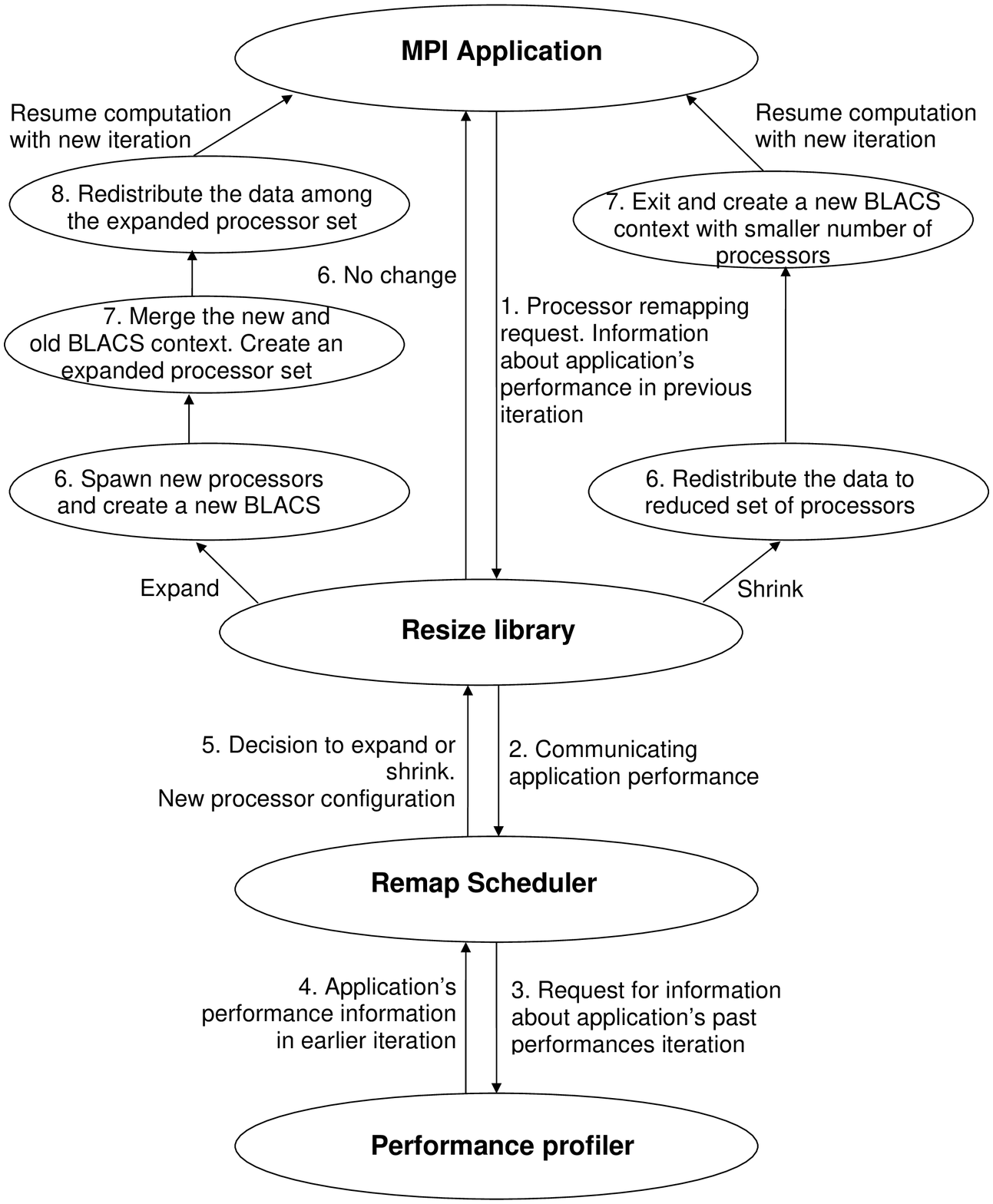}
\label{fig:statediagram}
}
\caption{(a) Architecture of \reshape{} (b) State diagram for app\-lication ex\-pansion and shrinking
}
\end{center}
\end{figure}

The \reshape{} framework, shown in Figure~\ref{fig:design}{}, consists of two main components. The first component is the application scheduling and monitoring module which schedules and monitors jobs and gathers performance data in order to make resizing decisions based on application performance, available system resources, resources allocated to other jobs in the system and jobs waiting in the queue.
The second component of the framework consists of a programming model for resizing applications.  This includes a resizing library and an API for applications to communicate with the scheduler to send performance data and actuate resizing decisions. The resizing library includes algorithms for mapping processor topologies and redistributing data from one processor topology to another. The individual components in these modules are explained in detail by Sudarsan and Ribbens~\cite{reportRajeshCorr}.

\subsection{Resizing library}
The resizing library provides routines for changing the size of the processor set assigned to an application and for mapping processors and data from one processor set to another. An application needs to be  re-compiled with the resize library to enable the scheduler to dynamically add or remove processors to/from the application. During resizing, rather than suspending the job, the application execution control is transferred to the resize library which maps the new set of processors to the application and  redistributes the data (if required). Once mapping is completed, the resizing library returns control back to the application and the application continues with its next iteration. The application user needs to indicate the global data structures and variables so that they can be redistributed to the new processor set after resizing. Figure~\ref{fig:statediagram}{} shows the different stages of execution required for changing the size of the processor set for an application.

Our API gives programmers a simple way to indicate \textit{resize points} in the application, typically at the end of each iteration of the outer loop. At resize points, the application contacts the scheduler and provides performance data to the scheduler.  The metric used to measure performance is the time taken to compute each iteration. The scheduler's decision to expand or shrink the application is passed as a return value.
If an application is allowed to expand to more processors, the response from the Remap Scheduler includes the size and the list of processors to which an application should expand.
A call to the redistribution routine remaps the global data to the new processor set. If the Scheduler asks an application to shrink, then the  application first redistributes its global data across a smaller processor set, retrieves its previously stored MPI communicator,  and creates a new BLACS~\cite{dongarra} context for the new processor set. The additional processes are terminated when the old BLACS context is exited.  The resizing library notifies the Remap Scheduler about the number of nodes relinquished by the application.

\subsection {Application Programming Interface (API)}
A simple API allows user codes to access the \reshape{} framework and library. The core functionality is accessed through the following internal and external interfaces.
These functions are available for use by advanced application programmers. These functions provide the main functionality of the resizing library by contacting the scheduler, remapping the processors after an expansion or a shrink, and redistributing the data. These functions are listed as follows:
\begin {itemize}
\item   \textit{reshape\_Initialize (global data array, nprocessors, blacs\_context, iterationCount, processor\_row, \-processor\_\-column, job\_id)}: initializes the iterationCount and the global data array with the initial values and creates  a blacs\_context for the two-dimensional processor topology. The function returns values for processor row, column configuration and job\_id.
\item    \textit{reshape\_ContactScheduler(iteration\_time, redistribution\_time, processor\_row\_}\textit{\-count, \-processor\-\_column\-\_count, \-job\_id)}:
 contacts the scheduler and supplies last iteration time; on  return, the scheduler indicates whether the application should expand, shrink, or continue execution with the current processor size.
\item    \textit{reshape\_Expand ()}: adds the new set of processors (defined by previous call to reshape\_contactScheduler) to the current set using BLACS.
\item    \textit{reshape\_Shrink ()}: reduces the processor set size (defined by previous call to reshape\_contactScheduler) to an earlier configuration and relinquishes additional processors.
\item   \textit{reshape\_Redistribute(Global data array, current BLACS context, current processor set size, EXPAND/SHRINK)}:  redistributes global data among the newly spawned or shrunk processors. The redistribution time is computed and stored for next resize point.
\item \textit{reshape\_Log (starttime, endtime)}: computes the average iteration time of the current iteration for all the processors and stores it for next resize point.
\end{itemize}

\begin{figure}[!ht]
\begin{center}
\subfigure[]{
\includegraphics[scale=0.45]{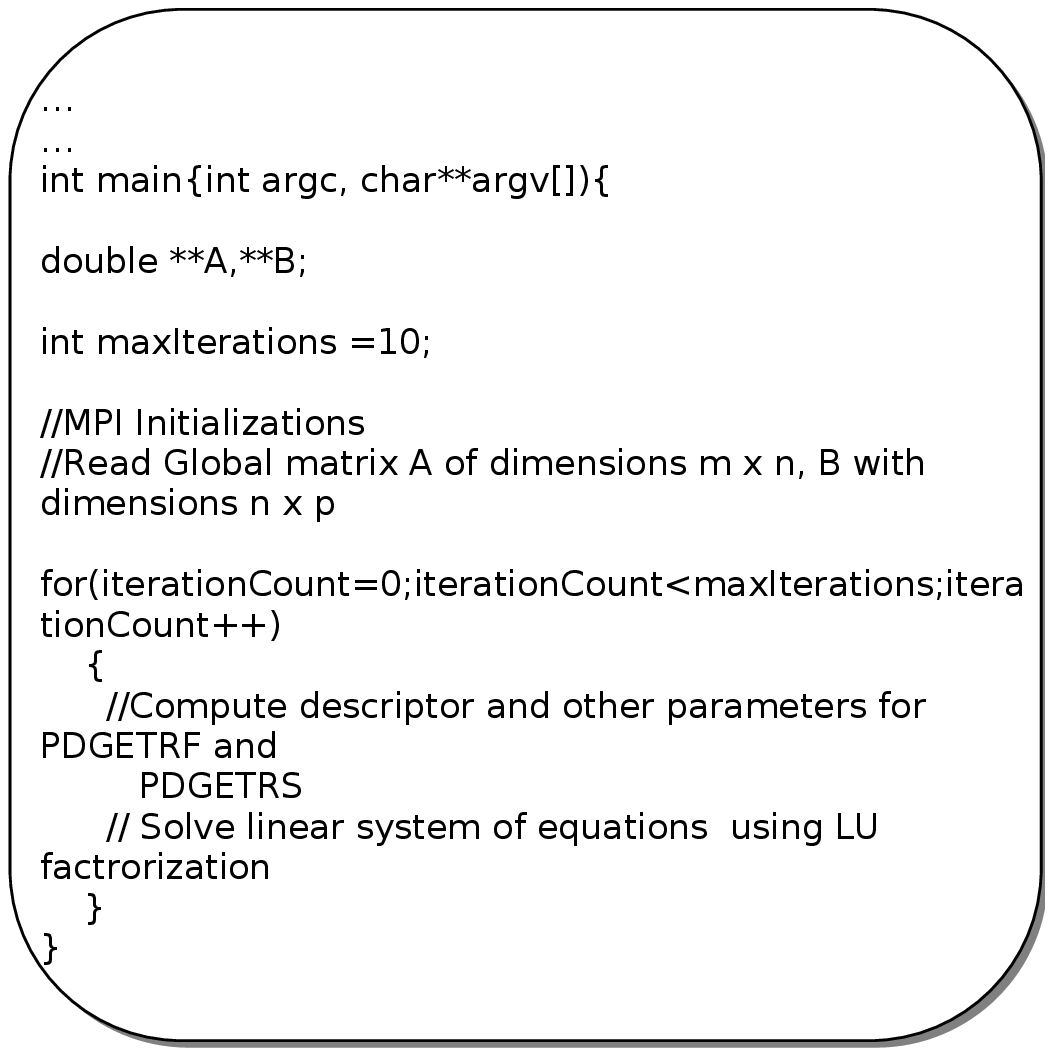}
\label{fig:originalcode}
}
\subfigure[]{
\includegraphics[scale=0.59]{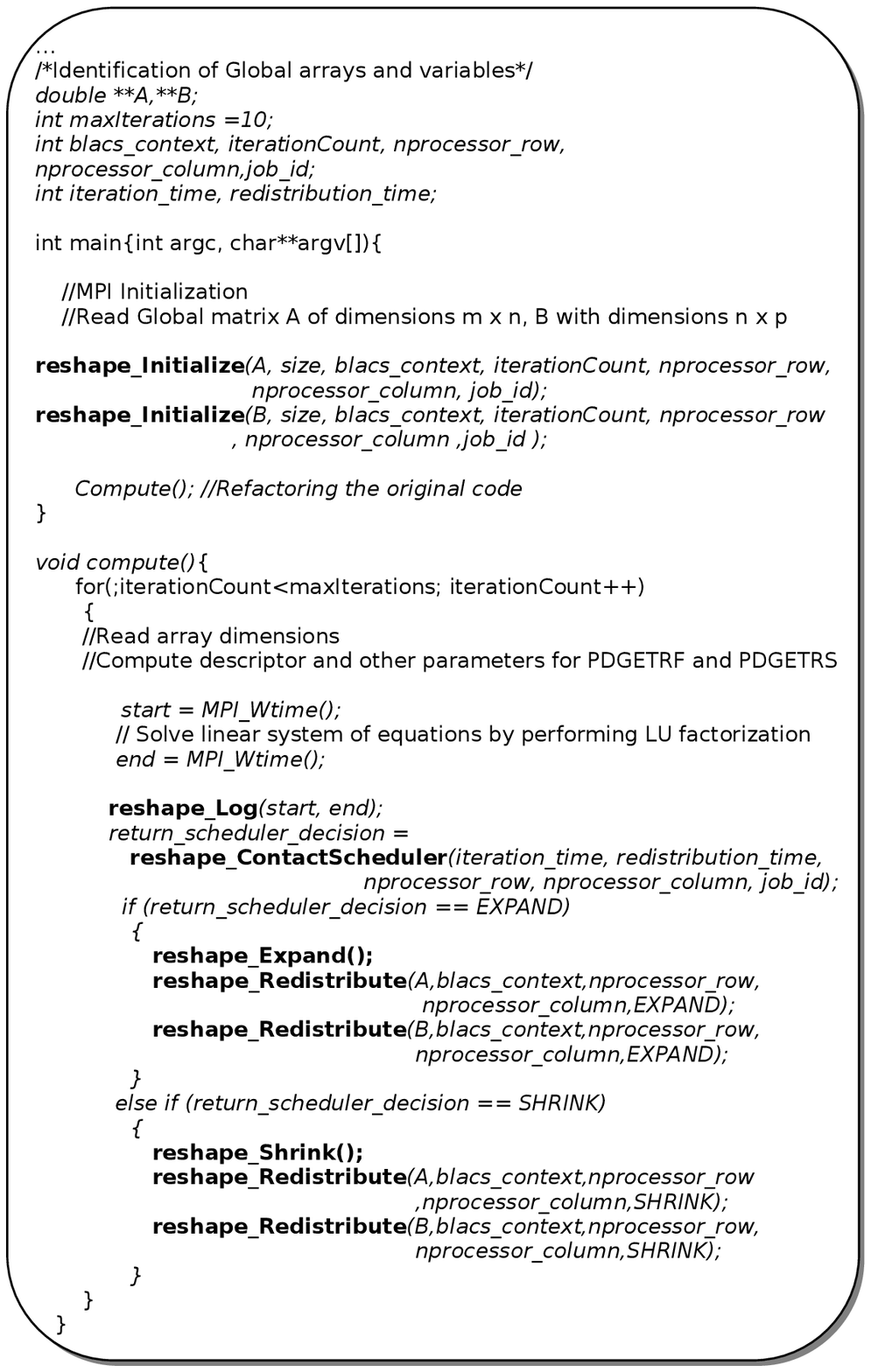}
\label{fig:modifiedcode}
}
\end{center}
\caption{(a) Original MPI code for solving system of linear equations. (b) Code modified for resizing using ReSHAPE's API}
\end{figure}

Figure~\ref{fig:originalcode} shows the source code for a simple MPI application for solving a sequence of linear system of equations using ScaLAPACK functions.
The original code was refactored to identify the global data structures and variables. The ReSHAPE API calls were inserted at the appropriate locations in the refactored code.
Figure~\ref{fig:modifiedcode} shows the modified code. 
\subsection{Data Redistribution}
\label{subsec:Dataredist}
The data redistribution library in ReSHAPE uses an efficient algorithm for redistributing block-cyclic  arrays between processor sets organized in a 1-D (row or column format) or checkerboard processor topology. 
The algorithm for redistributing 1-D block-cyclic array over a one-dimensional processor topology was first proposed by Park et al.~\cite{park}. 
We extend this idea to develop an algorithm to redistribute both one- and two-dimensional block-cyclic data 
across a two-dimensional processor grid of processors.
In our redistribution algorithm, we assume the following:
\begin{itemize}
\item 	Source processor configuration: $P_{r} \times P_{c}$ ($rows \times columns$), $P_r$,\, $P_c > 0$.
\item 	Destination processor configuration: $Q_{r} \times Q_{c}$ ($rows \times columns$), $Q_r$,\, $Q_c > 0$.
\item 	The data granularity is set at the block level, i.e., a block is the smallest data that will be transferred which cannot be further subdivided. 
        This block size is specified by the user.
\item 	The data matrix, \textit{data}, which needs to be redistributed, is of dimension $n \times n$.
\item 	Let the block size be \textit{NB}. Therefore total number of data blocks = $(n/NB)\ast (n/NB)$
      = $N \times N$, represented using matrix \textit{Mat}.
\item 	We use $Mat(x$,$\,y)$ to refer $block(x,y)$, $0 \leq x, y < N$.
\item   The data that can be equally divided among the source and destination processors P and Q respectively,
        i.e., $N$ is evenly divisible by $P_{r}$, $P_{c}$, $Q_{r}$, and $Q_{c}$. Each processor has an integer number of data blocks.
\item The source processors are numbered $P_{(i,j)}$, $0 \leq i < P_r$, $0 \leq j< P_c$ and the destination processors are numbered as $Q_{(i,j)}$, $0 \leq i < Q_r$, $0 \leq j<Q_c$
\end{itemize}

\subsubsection{Problem Definition.}
We define 2D block-cyclic distribution as follows: Given a two dimensional array of $n \times n$ elements with block size \textit{NB} 
and a set of $P$ processors arranged in checkerboard topology, the data is partitioned into $N \times N$ blocks 
and distributed across $P$  processors, where $N = n/NB$.
Using this distribution a matrix block, $Mat(x, y)$, is assigned to the source processor $P_{c} \ast (x\% P_{r})+y\%P_{c}$, $0 \leq x < N$, $0 \leq y< N$.
Here we study the problem of redistributing a two-dimensional block-cyclic matrix from $P$ processors to 
$Q$ processors arranged in checkerboard topology, where $P \neq Q$ and $NB$ is fixed. After redistribution, the block 
$Mat(x, y)$ will belong  to the destination processor $Q_{c} \ast (x\% Q_{r})+y\% Q_{c}$, $0 \leq x < N$, $0 \leq y < N$.

\begin{figure}[!ht]
\flushleft
\subfigure[]{
\includegraphics[scale=0.46]{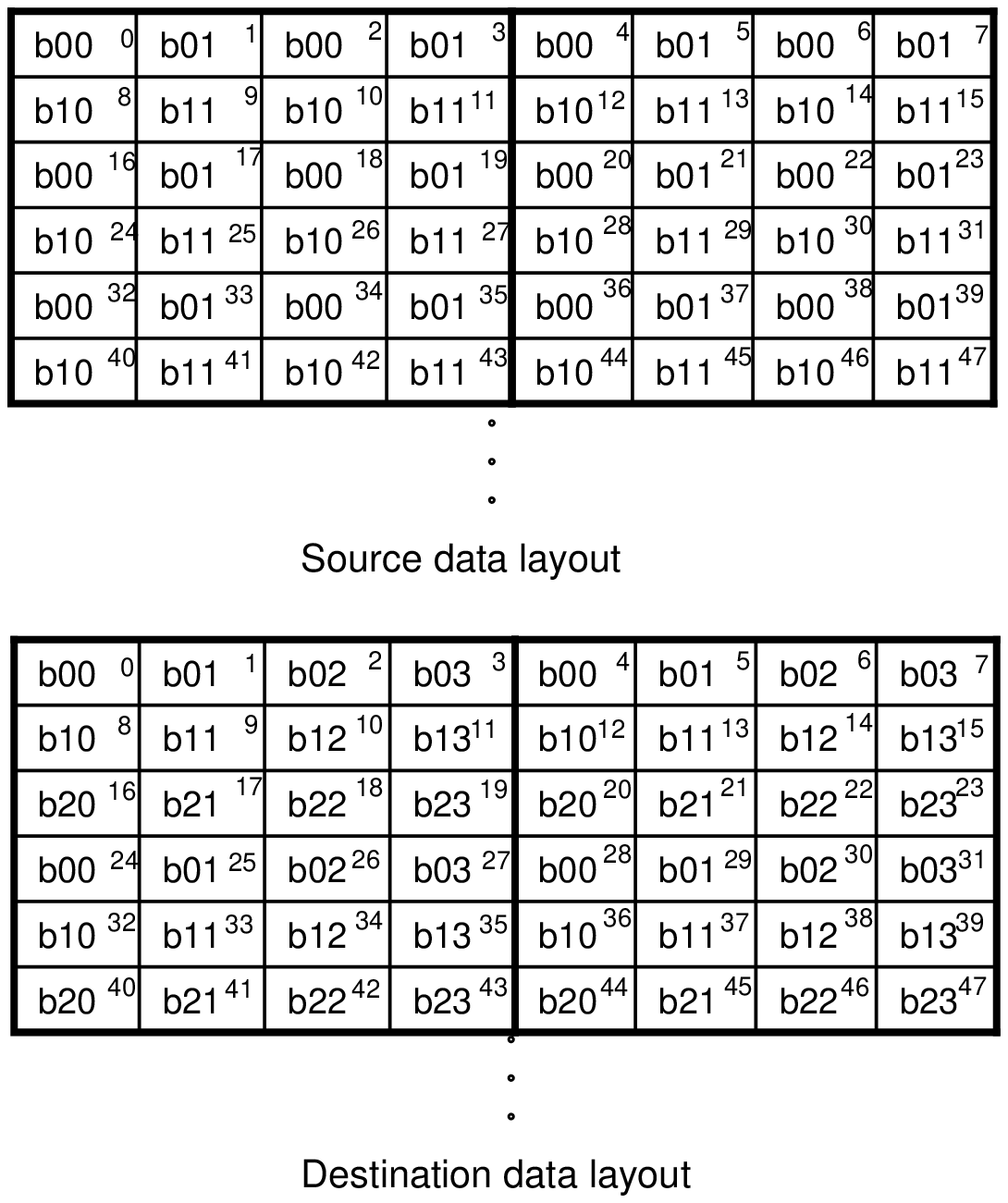}
\label{fig:datalayout}
}
\subfigure[]{
\includegraphics[scale=0.46]{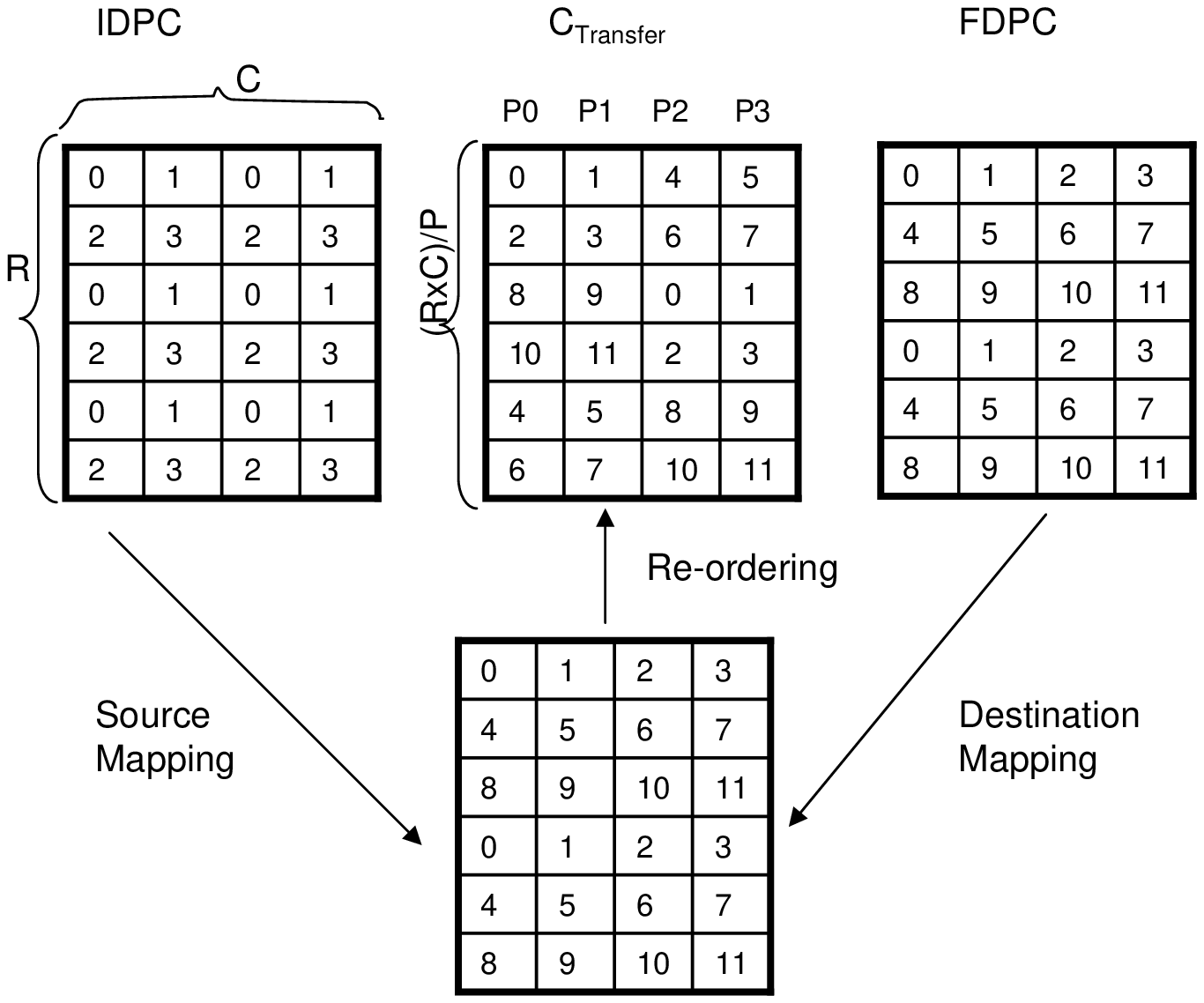}
\label{fig:redisttables}
}
\caption{(a) $P$ = $4$ ($2\times 2$), $Q$ = $12$ ($3 \times 4$)  Data layout in source and destination proce\-ssors. (b) Creating of Communication Schedule ($C_{Transfer}$) from
Initial Data Processor Configuration table (IDPC), Final Data Processor Configuration table (FDPC)}
\end{figure}

\subsubsection{Redistribution Terminologies.}
\begin{description}
\item[(a)] \textbf{Superblock}: Figure~\ref{fig:datalayout} shows the checkerboard distribution of a $8 \times 6$ block-cyclic data on source and destination processor grids. 
The $b00$ entry in the source layout table indicates that the block of data is owned by processor $P_{(0,0)}$, block denoted by $b01$ is owned by processor $P_{(0,1)}$ and so on. 
The numbers on the top right corner in every block indicates the id of that data block. From this data layout, a periodic pattern can be identified for redistributing
data from source to destination layout. The blocks $Mat(0,0)$, $Mat(0,2)$, $Mat(2,0)$, $Mat(2,2)$, $Mat(4,0)$ and $Mat(4,2)$, owned by processor $P_{(0,0)}$ in the source layout, are transferred to processors $Q_{(0,0)}$,
$Q_{(0,2)}$, $Q_{(2,0)}$, $Q_{(2,2)}$, $Q_{(1,0)}$ and $Q_{(1,2)}$. This mapping pattern repeats itself for blocks $Mat(0,4)$, $Mat(0,6)$, $Mat(2,4)$, $Mat(2,6)$, $Mat(4,4)$ and $Mat(4,6)$. 
Thus we can see that the communication pattern of the blocks $Mat(i, j)$, $0 \leq i< 5$, $0 \leq j < 4$ repeats for other blocks in the data. 
A superblock is defined as the  smallest set of data blocks whose mapping pattern 
from source to destination processor can be uniquely identified. 
For a 2-D processor topology data distribution, each superblock is represented as 
a table of R rows and C columns, where 

\hspace{0.2in}  $R = lcm(P_{r}$,\,$Q_{r})$\hspace{1in} $C = lcm(P_{c}$,\,$Q_{c})$

The entire data is divided into multiple superblocks and the mapping pattern of
the data in each superblock
is identical to the first superblock, i.e., the data blocks located at the same relative position                                                                               in all the superblocks are                                                              transferred to the destination processor.
A 2-D block matrix with $Sup$ elements is used to represent the entire data where each 
element is a Superblock. The dimensions of this block matrix are $Sup_R$ and $Sup_C$ where,

\hspace{0.2in}$Sup_R = N/R$ \hspace{0.5in} $Sup_C = N/C$ \hspace{0.5in}$Sup = (N/R\ast N/C)$
\item[(b)] \textbf{Layout}: Layout is an 1-D array of $Sup_R \ast Sup_C$ elements where each element 
is a 2-D table which stores the block ids present in that superblock. There are $Sup$ number of 2-D tables in the Layout array 
where each table has the dimension $R\times C$.

\item[(c)] \textbf{Initial Data-Processor Configuration (IDPC)}: This table represents the initial processor layout for the 
data before redistribution for a single superblock. Since the data-processor mapping is identical over all the superblocks, 
only one instance of this table is created. The table has $R$ rows $\times C$ columns.
$IDPC(i, j)$ contains the processor id $P_{(i, j)}$ that owns the block $Mat(i, j)$ located at the same relative position in all the superblocks, 
($0 \leq i < ,R$, $0 \leq j < C$).

\item[(d)] \textbf{Final Data-Processor Configuration (FDPC)}: The table represents the final processor configuration for 
the data layout after redistribution for a single super\-block. Like \textit{IDPC}, only one instance of this table is 
created and used for all the data superblocks. The dimensions of this table is $R \times C$. \textit{FDPC(i, j)} contains the processor 
id $Q_{(i, j)}$ that owns the block $Mat(i, j)$ after redistribution located at the same relative position in all the superblocks,
($0 \leq i < R$, $0 \leq j< C$).

\item[(e)]The source processor for any data block \textit{Mat(i, j)} in the data matrix can be computed using the formula

\hspace{1.3in}$Source(i,j) = P_{c} \ast(i\% P_{r}) +(j\% P_{c})$

\item[(f)] \textbf{Communication schedule send table ($C_{Transfer}$)}: This table contains the final communication schedule for redistributing 
data from source to destination layout.  This table is created by re-ordering the \textit{FDPC} table. 
The columns of $C_{Transfer}$ correspond to $P$ source processors and the rows correspond to individual 
communication steps in the schedule.
The number of rows in this table is determined by $(R \ast C)/P$. 
The network bandwidth is completely utilized in every communication step as the schedule
involves all the source processors in data transfer. A positive entry in the $C_{Transfer}$ table
indicates that in the $i^{th}$ communication step, processor $j$ will send data to $C_{Transfer}(i,j)$, 
$0 \leq i < (R \ast C)/P$, $0 \leq j < (P_r \ast P_c)$.

\item[(g)] \textbf{Communication schedule receive table ($C_{Recv}$)}: This table is derived from the $C_{Transfer}$ table where the 
columns correspond to the destination processors. The table has the same number of rows as the $C_{Transfer}$ table.  
A positive entry at $C_{Recv}(i,j)$ indicates that processor $j$ will receive data 
from source processor at $C_{Recv}(i,j)$ in the $i^{th}$ communication step, $0 \leq i< (R \ast C)/P$, $0 \leq j < (Q_r \ast Q_c)$. 
If $(Q_{r} \ast Q_{c}) \ge (P_{r} \ast P_{c})$, 
then the additional entries in the $C_{Recv}$ table are filled with -1.

\end{description}

\subsubsection{Algorithm.}
\begin{description}
\item[\textbf{Step 1:}] \hspace{0.05in}\textit{Create Layout table}\\
The Layout array of tables are created by traversing through all the data blocks in 
matrix $Mat(i, j)$, where $0\leq i$, $j < N$, $0 \leq j <N$. The superblocks in $Mat(i, j)$ is traversed in row-major format.  

\textbf{Pseudocode:}
\vspace{-0.1in}
\begin{tabbing}
\hspace{0.3in}\textbf{for} $superblockcount \leftarrow 0$ \textbf{to} $Sup -1$ \textbf{do}\\ 
\hspace{0.5in}   \textbf{for}  $i \leftarrow 0$ \textbf{to} $R/P_{r} -1$ \textbf{do} \\
\hspace{0.7in}    \textbf{for}  $j \leftarrow 0$ \textbf{to} $C/P_{c} - 1$ \textbf{do}\\
\hspace{0.9in}     \textbf{for}  $k \leftarrow 0$ \textbf{to} $P_{r} - 1$ \textbf{do}\\
\hspace{1.1in}      \textbf{for}  $l \leftarrow 0$ \textbf{to} $P_{c} - 1 $ \textbf{do}\\ 
\hspace{1.3in}           $Layout[superblockcount](i\ast C/P_{c} + k, j \ast R/P_{r}+l) = $ \\
\hspace{2.2in}            $Mat(superblockid_{row}\ast R + i\ast P_{c} + k,$ \\
\hspace{2.2in}   	      $superblockid_{col}\ast C + j\ast P_r + l)$ \\

\hspace{0.5in} \textbf{if}$(reached\ end\ of\ column)$ \textbf{then}\\
\hspace{0.7in}	$increment\ Sup_R$\\
\hspace{0.7in}	$Sup_C \leftarrow 0$\\
\hspace{0.5in}\textbf{else}\\
\hspace{0.7in}	$increment\ Sup_C$\\
\end{tabbing}
\vspace{-0.2in}
\item[\textbf{Step 2:}] \hspace{0.05in}\textit{Creating IDPC and FDPC tables}\\
An entry at $IDPC (i$, $j)$  is calculated using the index $i$ and $j$ of the table and 
the size of the source processor set $P$, $0\leq i<R$, $0\leq j<C$. 
The Source function returns the processor id of the owner of the data before redistribution stored in that location.

Similarly, an entry $FDPC (i$, $j)$  is computed using the $i$ and $j$ coordinates of the table and the size of the destination processor set $Q$, $0\leq i<R$, $0\leq j<C$. The Source function returns the processor id of the 
owner of the redistributed data stored in that location.

\textbf{Pseudocode:}
\begin{tabbing}
\hspace{0.4in}\textbf{for} $i \leftarrow 0$ \textbf{to} $R - 1$ \textbf{do}\\
\hspace{0.7in}  \textbf{for} $j \leftarrow 0$ \textbf{to} $C -1 $ \textbf{do} \\
\hspace{1in}      $IDPC(i, j) \leftarrow  Source(i,j) \leftarrow P_{c}\ast (i\% P_{r}, j\% P_{c})$ \\
\\
\\
\hspace{0.4in}\textbf{for} $i \leftarrow 0$ \textbf{to} $R - 1$ \textbf{do} \\
\hspace{0.7in}   \textbf{for} $j \leftarrow 0$ \textbf{to} $C - 1 $ \textbf{do}\\
\hspace{1in}      $FDPC(i, j) \leftarrow  Source(i,j) \leftarrow  Q_{c} \ast (i\% Q_{r}, j\% Q_{c})$
\end{tabbing}
\item[\textbf{Step 3:}] \hspace{0.05in}\textit{Communication schedule tables($C_{Transfer}$ and $C_{Recv}$)}\\
The $C_{Transfer}$ table stores the final communication schedule for transferring data between 
the source and the destination processors. The columns in $C_{Transfer}$ correspond to source processor $P_{(i,j)}$. 
The table has $C_{TransferRows}$ rows and ($P_{r}$ $\ast$ $P_{c}$) columns, where

\hspace{1.3in}$C_{TransferRows} = (R\ast C)/(P_{r}\ast P_{c})$ 

Each entry in the $C_{Transfer}$ table is filled by sequentially traversing the \textit{FDPC} table in row-major format. 
The data corresponding to each processor  inserted at the appropriate column at the next available location. 
An integer counter updates itself and keeps track of the next available location (next row) for each processor.

\textbf{Pseudocode:}
\begin{tabbing}
\hspace{0.4in}$processor\_id = IDPC(i, j)$ \\
\hspace{0.4in}$C_{Transfer}(counter_{j}, processor\_id) \leftarrow FDPC(i,j)$ \\
\hspace{0.4in}$Update\ counter_{j}$
\end{tabbing}

where $0\leq i < R$ and $0 \leq j < C$.
Each row in the $C_{Transfer}$ table forms a single communication step where all the source processors send the 
data to a unique destination processor. The $C_{Recv}$ table is used by the destination processors to know the 
source of their data in a particular communication step.

\hspace{1.3in}{$C_{Recv}(i, C_{Transfer}(i,j)) = j$}

where $0 \leq i< C_{TransferRows}$ and  $0 \leq j < (Q_{r} \times Q_{c})$.

Node contention can occur in the $C_{Transfer}$ communication schedule if any one of the following conditions are true\\
(i)   $P_{r} \ge Q_{r}$ \\
(ii)  $P_{c} \ge Q_{c}$ \\
(iii) $P_{r} \ge Q_{r}$ and $P_{c} \ge Q_{c}$\\

If there are node contentions in the communication schedule, create a \textit{Processor Mapping} (PM) table  of dimension $R\times C$ and initialize it with the values from FDPC table.
To reduce node contentions, the \textit{PM} tables are circularly shifted in row or columns. To maintain data consistency, same operations 
are performed on the IDPC table and the superblock tables within the Layout array.
The $C_{Transfer}$ table is created from the modified PM table.
We identify 3 situations where node contentions can occur.
Case 1 and case 2 are applicable during both expansion and shrinking of an application 
while Case 3 can occur only when an application is shrinking to a smaller  destination processor set.

Do the following operation on IDPC, PM and on each 2-D table in the Layout array.\\
\noindent
\textit{Case 1}: If $P_{r} > Q_{r}$ and $P_{c}<Q_{c}$ then
\begin{enumerate}
\item Create $(R/ P_{r})$ groups with $P_{r}$ rows in each group.
\item For $1\leq i<P_r$, perform a circular right shift on each row i by $P_{c} \ast i$ elements in each group.
\item Create the $C_{Transfer}$ table from the resulting \textit{PM} table.
\end{enumerate}

\noindent
\textit{Case 2}: If $P_{r} < Q_{r}$ and $P_{c}>Q_{c}$ then
\begin{enumerate}
\item Create $(C/P_{c})$ groups with $P_{c}$ columns in each group.
\item For $1\leq j<P_c$, perform a circular down shift on each column j by $P_{r} \ast j$ elements in each group.
\item Create the $C_{Transfer}$ table from the resulting \textit{PM} table.
\end{enumerate}

\noindent
\textit{Case 3}: If $P_{r} > Q_{r}$ and $P_{c} > Q_{c}$ then 
\begin{enumerate}
\item Create ($C/ P_{c}$) groups with $P_{c}$ columns in each group.
\item For $1\leq j<P_c$, perform a circular down shift each column j by $P_{r} \ast j$ elements in each group.
\item Create ($R/ P_{r}$) groups with $P_{r}$ rows in each group.
\item For $1\leq i<P_r$, perform a circular right shift each row i by $P_{c} \ast i$ elements in each group
\item Create the $C_{Transfer}$ table from the resulting \textit{PM} table.
\end{enumerate}

The $C_{Recv}$ table is not used when the schedule is not contention-free. Node contention results in  overlapping entries in the $C_{Recv}$ table 
thus rendering it as unusable.

\item[\textbf{Step 4:}] \hspace{0.05in}\textit{Data marshalling and unmarshalling}\\
If a processor's rank equal the value at $IDPC(i$, $j)$, then  
the processor collects the data from the relative indexes of all the superblocks in the Layout array. Each collection of data
over all the superblocks forms a single message for communication for processor j.

If there are no node contentions in the schedule, each source processor stores 
$(R \ast C)/(P_{r} \ast P_{c})$ messages, each of size $(N \ast N/(R \ast C))$ in the original order of the data layout.
The messages received on the destination processor are unpacked into individual blocks
 and stored 
at an offset of ($R/Q_r$) $\ast$ ($C/Q_c$) elements from the previous data block 
in the local array. The first data block is stored at $zero^{th}$ location of the 
local array.
If the communication schedule has  node contentions, the order of the messages are shuffled according to row or column transformations. In such cases, the destination processor performs reverse index computation and stores the data at the correct offset.

\item[\textbf{Step 5:}] \hspace{0.05in}\textit{Data Transfer}\\
The message size in each send communication is equal to $(N \ast N)/(R\ast C)$ data blocks.
Each row in the $C_{Transfer}$ table corresponds to  a single communication step. 
In each communication step, the total volume of messages exchanged between the processors is $P \ast (N\ast N/(R\ast C))$ data blocks. 
This volume includes cases where data is locally copied to a processor without performing a MPI\_Send and MPI\_Recv operation.
 In a single communication step j, a source processor $P_i$ sends the marshalled  
message to the destination processor given by $C_{Transfer}(j,i)$, where  $0\leq j<C_{TransferRows}$, $0\leq i<(P_r \ast P_c)$, 
 
\paragraph*{Data Transfer Cost. }
For every communication call using MPI\_Send and MPI\_Recv, there is a latency overhead associated with it. Let us denote this time to initiate a message by $\lambda$. 
Let $\tau$ denote the time taken to transmit a unit size of message from source to destination processor. Thus, the time taken to send a message from a source processor
in single communication step is $((N \ast N)/(R\ast C))\ \ast \tau$. The total data transfer cost for redistributing the data across destination processors is
$C_{TransferRows} \ast (\lambda +((N \ast N)/(R\ast C)) \ast \tau)$.
\end{description}

\section{Experiments and Results}
\label{sec:experiments}
This section presents experimental results which demonstrate the performance of our two-dimensional block-cyclic redistribution algorithm.
The experiments were conduct\-ed on 50 nodes of a large  homogeneous cluster (System X).  Each node is a dual 2.3 GHz PowerPC 970 processor with 4GB of main memory.
Message passing was done using MPICH2~\cite{mpich2} over a Gigabit Ethernet
interconnection network. We integrated the redistribution algorithm  into the resizing library and evaluated its performance by measuring the total time taken by the algorithm to redistribute block-cyclic matrices  from P to Q processors. 
We present results from two sets of experiments. The first set of experiments evaluates the performance of the algorithm for resizing and compares it with 
the Caterpillar algorithm.
The second set of experiments focuses on the effects of processor topology on the redistribution cost. Table~\ref{tab:procconfig} shows all 
the possible processor configurations for various processor topologies. Processor configurations for the one-dimensional processor topology ($1\times Q_r \ast Q_c$ or $Q_r\ast Q_c \times 1$)  are
not shown in the table. For the two set of experiments described in this section, we have used the following matrix sizes - $2000 \times 2000, 4000 \times 4000, 6000 \times 6000,
8000 \times 8000, 12000 \times 12000, 16000 \times 16000, 20000 \times 20000$ and $24000\times 24000$. 
A problem size of $8000$ indicates the matrix $8000\times 8000$. 
The processor configurations listed in Table~\ref{tab:procconfig} evenly
divide the problem sizes listed above. 

\begin{table}[ht]
\begin{center}
\caption{Processor configuration for various topologies}
\label{tab:procconfig}
\begin{tabular}{|l|l|}
\hline
Topology&Processor configurations \\
\hline
Nearly-square&$1\times2$, $2\times2$, $2\times3$, $2\times4$, $3\times3$, $3\times4$, $4\times4$, $4\times5$, $5\times5$,\\
&$5\times6$, $6\times6$, $5\times8$, $6\times8$\\
\hline
Skewed-rectangular &$1\times2$, $2\times2$, $2\times3$, $2\times4$, $3\times3$, $2\times6$, $2\times8$, $2\times10$, $5\times5$,\\
&$3\times10$, $2\times18$, $2\times20$, $2\times24$, $2\times1$, $3\times2$, $4\times2$, $6\times2$, \\
&$8\times2$, $10\times2$, $10\times3$, $18\times2$, $20\times2$, $24\times2$\\
\hline
\end{tabular}
\end{center}
\end{table}

\subsection{Overall Redistribution Time}
\begin{figure}[ht]
\subfigure[Expansion]{
\includegraphics[scale=0.46]{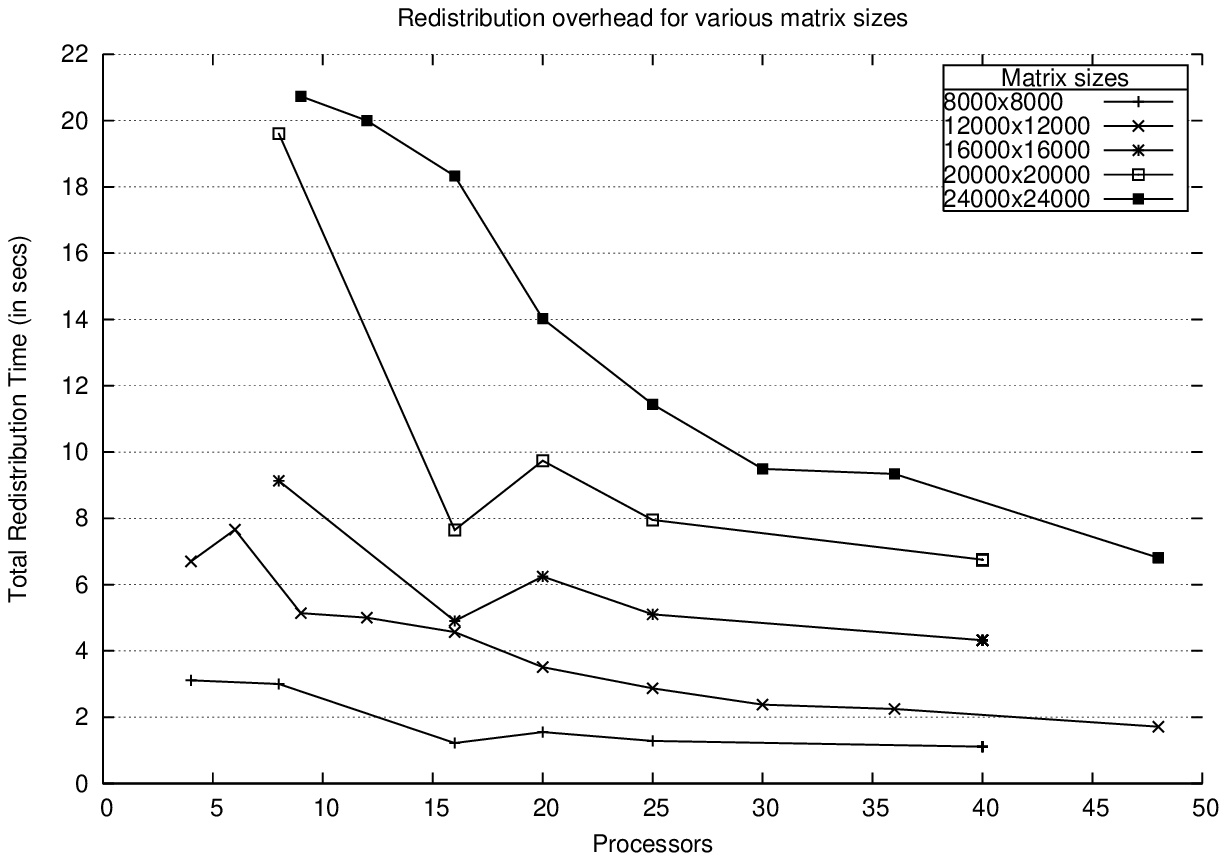}
\label{fig:redistribution}
}
\subfigure[Shrinking]{
\includegraphics[scale=0.46]{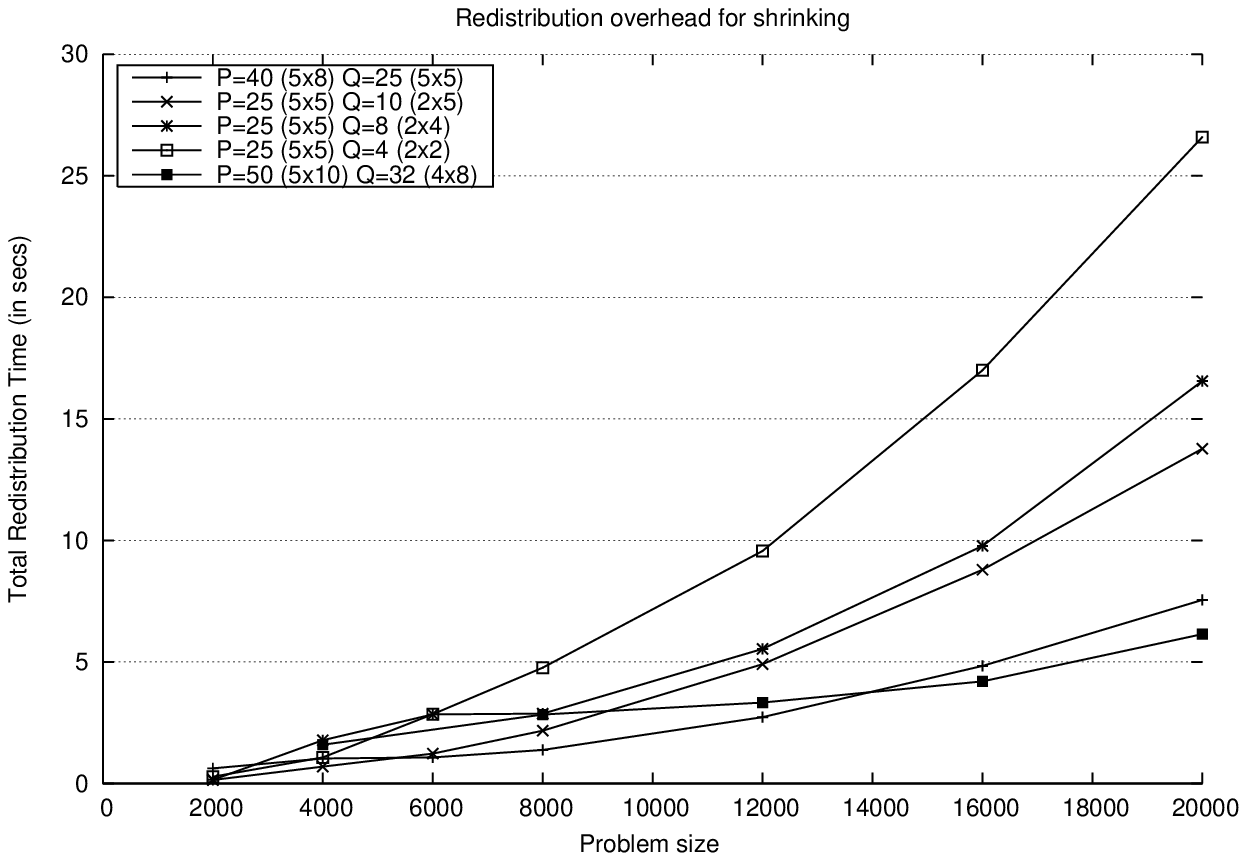}
\label{fig:shrinking}
}
\caption{Redistribution overhead incurred while resizing using ReSHAPE.}
\end{figure}

Every time an application acquires or releases processors, the globally
distributed data has to be redistributed to the new processor topology.
Thus, the application incurs a redistribution overhead each time it expands or shrinks.
We assume a nearly-square processor topology for all the  processor sizes used in this experiment.
The matrix stores data as double precision floating point numbers.
Figure~\ref{fig:redistribution} shows the overhead for
redistributing large dense matrices
for different matrix sizes using the our redistribution algorithm.
Each data point in the graph represents the data redistribution cost
incurred when increasing the size of the processor configuration from the
previous (smaller) configuration. Problem size 8000 and 12000 start execution with 2 processors,
problem size  16000 and 20000 start with 4 processors, and the 24000
case starts with 6 processors. The starting processor size is the smallest size which can accommodate the data. 
The trend shows that the redistribution cost increases with matrix size,
but for a fixed matrix size
the cost decreases as we increase the number of processors. 
This makes sense because for small processor size, the amount of data per processor that must be transferred is large.
Also the communication schedule developed
by our redistribution algorithm is independent of the problem size and depends only on the  source and destination processor set size.

\vspace{-0.02in}
Figure~\ref{fig:shrinking} shows the overhead cost incurred while shrinking large matrices from $P$ processors to $Q$ processors. In this 
experiment, we assign the values for $P$ from the set $25$, $40$, $50$ and $Q$ from the set $4$, $8$, $10$, $25$  and $32$.
Each data point in the graph 
represents the redistribution overhead incurred while shrinking at that problem size. From the graph, it is evident that the redistribution 
cost increases as we increase the problem size. Typically, a large difference between the 
source and destination processor set results in higher redistribution cost. The rate at which the redistribution cost increases  depends on the size 
of source and destination processor set. But we note that  smaller destination processor set size 
has a greater impact on the redistribution cost compared to the difference between the processor set sizes.
This is shown in the graph where the redistribution cost for shrinking from $P=50$ to $Q=32$ is lower compared to the cost  
when shrinking from $P=25$ to $Q=10$ or $P=25$ to $Q=8$.
\begin{figure}[th]
\subfigure[Redistribution overhead while resizing from 8 to 40 processors]{
\includegraphics[scale=0.46]{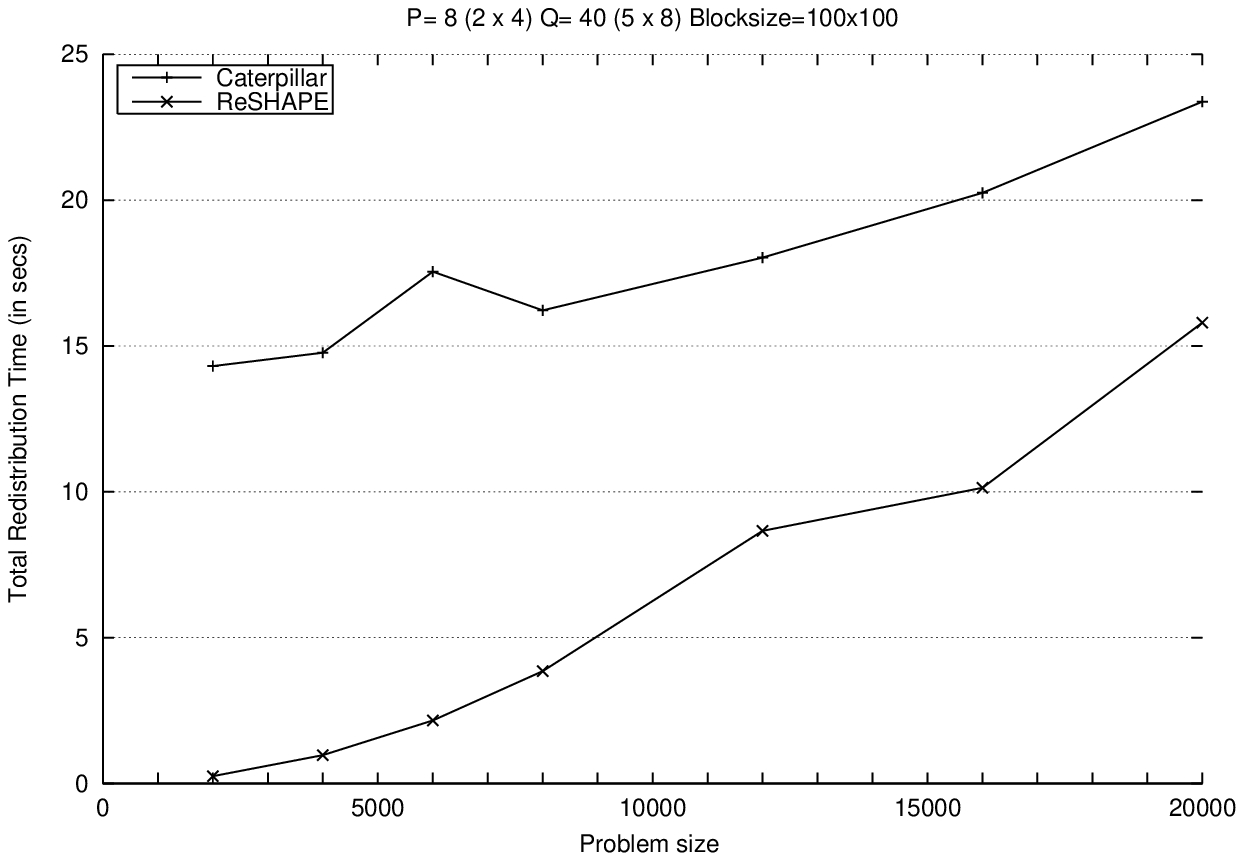}
\label{fig:expand8to40}
}
\subfigure[Redistribution overhead while resizing from 8 to 50 processors]{
\includegraphics[scale=0.46]{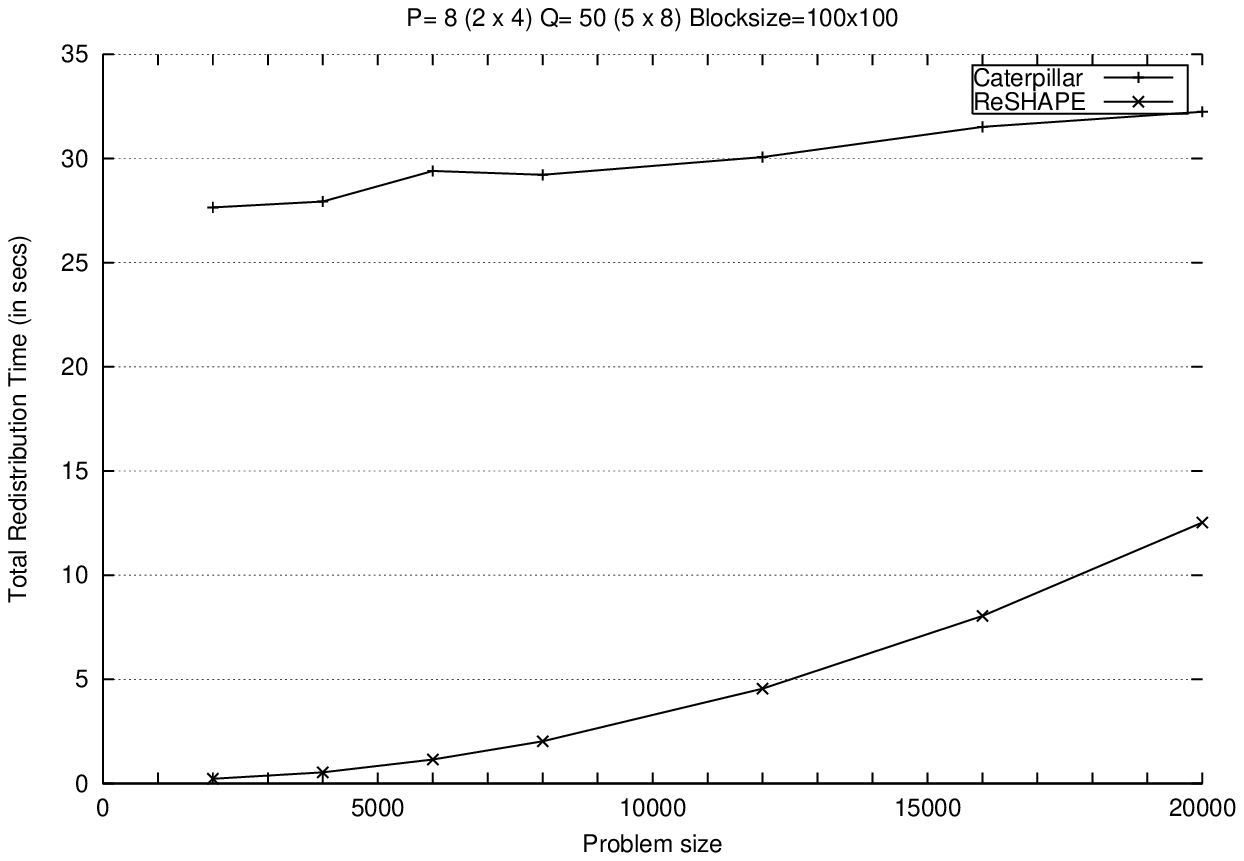}
\label{fig:expand8to50}
}
\caption{Comparing the total redistribution time for data redistribution in our algorithm with  Caterpillar algorithm}
\end{figure}

\vspace{-0.02in}
Figure~\ref{fig:expand8to40} and \ref{fig:expand8to50} compares  the total redistribution cost of our algorithm and the Caterpillar algorithm. 
We have not compared the redistribution costs with the bipartite redistribution algorithm as our algorithm assumes that data redistribution from 
P to Q processors includes an overlapping set processors from the 
source and destination processor set. 
The total redistribution time is the sum total of schedule computation time, index computation time, 
packing and unpacking the data and the data transfer time.
 In each communication step, each sender packs a message before sending it and the receiver unpacks
the message after receiving it. The Caterpillar algorithm does not attempt to schedule communication operations and send equal sized messages in each step. Figure~\ref{fig:expand8to40} shows experimental results for redistributing block-cyclic two-dimensional arrays from a $2 \times 4$ processor 
grid to a $5 \times 8$ processor grid. On average, the total redistribution time of our algorithm is 12.7 times less than the Caterpillar algorithm. 
In Figure~\ref{fig:expand8to50}, the total redistribution time of our algorithm is about 32 times less than  of the Caterpillar algorithm. 
In our algorithm, the total number of communication calls for redistributing from 8 to 40 processors is 80 whereas in Caterpillar the number is 160. Similarly, the number of MPI communication calls in our algorithm for redistributing 2D block-cyclic array from 8 processors to 50 processors is 196 as compared to 392 calls in the Caterpillar algorithm.

\subsection{Effects of Processor Topology on Total Redistribution Time}
\begin{figure}[ht]
\subfigure[$20000 \times 2000$ matrix with different topologies]{
\includegraphics[scale=0.46]{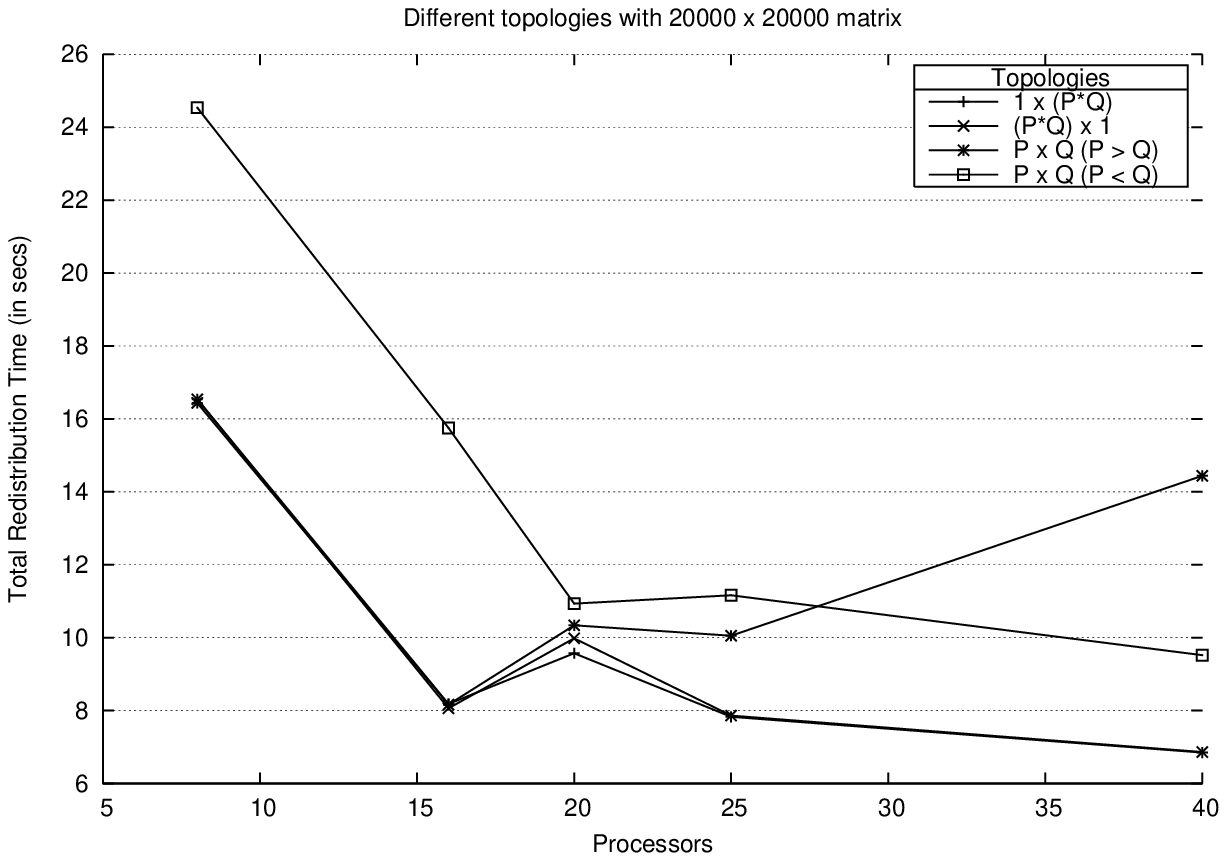}
\label{fig:topology20000}
}
\subfigure[$24000 \times 2400$ matrix with different topologies]{
\includegraphics[scale=0.46]{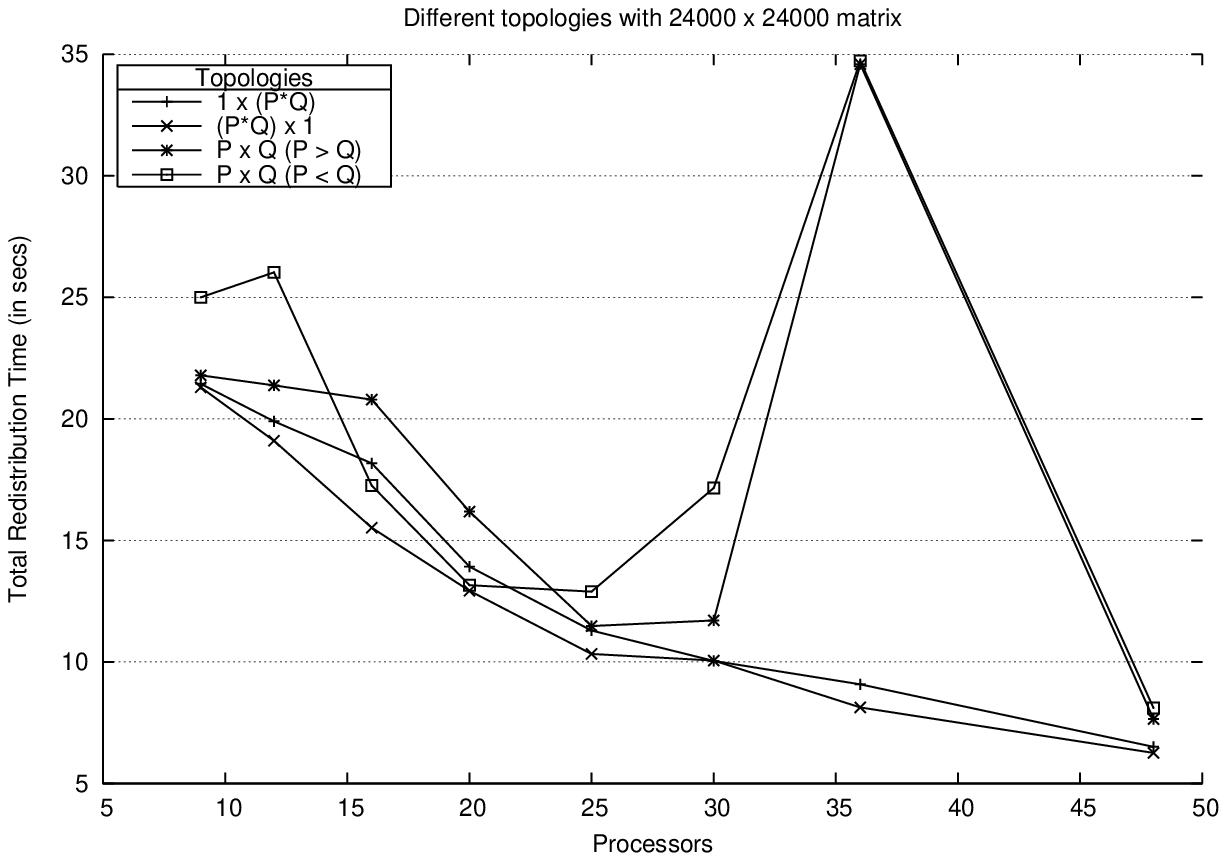}
\label{fig:topology24000}
}
\caption{Effects of skewed processor topologies on total redistribution time}
\end{figure}

\vspace{-0.02in}
In this experiment, we report the performance of our redistribution algorithm with four different processor topologies --- One-dimensional-row (Row-major), 
One-\-dim\-ensional-column (Column major), Skewed-rectangular-row ($P_r \times P_c$, $P_r >Pc$) and 
Skewed-rectangular-column ($P_r \times P_c$, $P_r < P_c$).
 The processor configurations used for the Skewed-rectangular topologies are listed in Table~\ref{tab:procconfig}.
Figure~\ref{fig:topology20000} and Figure~\ref{fig:topology24000} shows the overhead for
redistributing problem size 20000 and 24000
across different processor topologies using the our redistribution algorithm, respectively.
The total redistribution cost for redistributing $20000 \times 20000$ matrix across an one-dimensional topology is  comparable to the total redistribution cost on a nearly-square processor topology (see Figure~\ref{fig:redistribution}). 
In the case of skewed-rectangular topologies,  the  total redistribution time is slightly higher compared to the redistribution cost with nearly-square processor topologies. We ran this experiment on other problem sizes --- $8000 \times 8000$ and $16000 \times  16000$ and observed results similar to Figure~\ref{fig:topology20000}.
An increase in  the total redistribution time for skewed-rectangular topology can be due to one of the two situations. \\
(1) There is an increase in the total number of messages to be transferred using the communication schedule. \\
(2) Node contention in the communication schedule is high.\\ 

\vspace{-0.03in}
Since the dimensions of a superblock depends upon source and destination processor row and columns, a change in the processor topology can change the number of elements in a superblock. As a result, the number of messages exchanged between processors will also vary thereby increasing or decreasing
the total redistribution time. 
Figure~\ref{fig:topology24000} shows that the total redistribution cost for a skewed processor topology suddenly increases 
when the processor size increases from 30 to 36 ($10 \times 3$ to $18 \times 2$).
In this case the number of elements in superblock increases to 540.
Table~\ref{tab:topologycount}  shows the total MPI send/receive counts for redistributing between different processor sets on different topologies. 
From Table~\ref{tab:topologycount},
we note that data redistribution using a skewed-rectangular processor topology 
requires exactly half the number of send/receive operation as compared to 
nearly-square 
topology. The algorithm uses only 18 MPI send/receive operations to 
redistribute data from $4$ to  $20$ processors and 36 to redistribute from 
$8$ to $40$ processors as compared to 36 and 72 respectively required 
for a nearly-square topology.
In Figure~\ref{fig:topology20000}, the cost of redistribution in a  $P < Q$ topology is more than the redistribution cost for a $P > Q$ topology. The reason for this additional overhead can be attributed to increased number of node contentions in the comunication schedule for the $P < Q$ topology.  
The node contentions reduces as the processor size increases and the topology is maintained in subsequent iterations. When data is redistributed from $P$ = $25$ (square topology) to $Q$ = $40$ (skewed topology), 
 node contentions in the communication schedule of $Q$ = $40$ ($10\times 4$) are higher compared to the schedule for redistribution to $Q$ = $40$ ($4\times 10$).
\begin{table}[ht]
\begin{center}
\caption{ Counting topology dependent Send/Recvs. (P, Q) = size of source and destination processor set}
\label{tab:topologycount}
\begin{tabular}{|c|c|c|c|c|c|c|c|}
\hline
Redistribution&Communication&\multicolumn{2}{|c|}{Nearly square}&\multicolumn{2}{|c|}{1 Dimensional}&\multicolumn{2}{|c|}{Skewed-rectangle}\\
configuration&steps&Copy&Send/Recv&Copy&Send/Recv&Copy&Send/Recv\\
\hline
(2, 4)&2&2&2&2&2&2&2\\
\hline
(4, 6)&3&3&9&4&8&3&9\\
\hline
(4, 8)&2&2&6&4&4&2&6\\
\hline
(6, 9)&3&6&12&6&12&3&15\\
\hline
(8, 16)&2&8&8&8&8&4&12\\
\hline
(9, 12)&4&6&30&9&27&3&33\\
\hline
(12, 16)&4&12&36&12&36&12&36\\
\hline
(16, 20)&5&10&70&16&64&16&64\\
\hline
(20, 25)&5&20&80&20&80&5&95\\
\hline
(25, 30)&6&15&135&25&125&4&146\\
\hline
(25, 40)&8&7&193&20&180&25&175\\
\hline
(30, 36)&6&30&150&30&150&15&525\\
\hline
(36, 48)&4&12&132&36&108&36&108\\
\hline
(4, 20)&10, 5 (skewed)&2&38&4&36&2&18\\
\hline
(8, 40)&10, 5 (skewed)&8&72&8&72&4&36\\
\hline
(8, 50)&25&8&192&8&192&8&192\\
\hline
\end{tabular}
\end{center}
\end{table}

\section{Discussion and Future Work}
\label{sec:discussion}
In this paper we have introduced a framework, ReSHAPE,  that enables parallel message passing
applications to be resized during execution. 
We have extended the functionality of the resizing library in ReSHAPE to support redistribution of 2-D block-cyclic matrices distributed across a 2-D processor topology. 
We build upon the work by Park et al.~\cite{park} to derive an efficient 2-D redistribution algorithm. Our algorithm redistributes a two-dimensional block-cyclic data
distribution on a 2-D grid of P ($P_r \times P_c$) processors to  two-dimensional block-cyclic data distribution
on a 2-D grid with Q ($Q_r \times Q_c$) processors, where P and Q can be any  arbitrary positive value. 
The algorithm  ensures a contention-free communication schedule if  $P_r \leq Q_r, P_c \leq Q_c$. 
For all other conditions involving $P_r$, $P_c$, $Q_r$, $Q_c$, the  algorithm minimizes node contention in the communication schedule by performing a sequence of row or column circular shifts.
We also show the ease of use of API provided by the framework to port and execute applications to make use of ReSHAPE's dynamic resizing capability.
Currently the algorithm can redistribute $N \times N$ blocks of data on P processors to Q processors only if $Q_r$ and $Q_c$ evenly divide N so that all the processors have equal number of integer blocks. We plan to generalize this assumption so that the algorithm can redistribute data between P and Q processors for any arbitrary value of P and Q. 

We are currently evaluating \reshape{} framework with
different scheduling strategies for processor reallocation,
quality-of-service and advanced reservation services. We are also 
working towards adding resizing capabilities to several production scientific codes and
adding support for a wider array of distributed data structures and other data
redistribution algorithms.
Finally, we plan to make \reshape{} a more extensible framework so
that support for heterogeneous clusters, grid infrastructure, shared memory
architectures, and distributed memory architectures can be implemented  as
individual plug-ins to the framework.

\bibliographystyle{splncs}
\bibliography{reference}
\end{document}